\documentclass[12pt]{article}
\usepackage{epsfig}
\usepackage{cite}

\begin{document}

\title{On Estimate of Real Accuracy of EOP Prediction\footnote{In: ASP Conf. Ser., v.~208, Polar Motion: Historical and Scientific
  Problems, Proc. IAU Coll. 178, Cagliari, Italy, Sep 27-30, 1999, Eds. S. Dick, D. McCarthy, B. Luzum, 2000, 505-510.}}
\author{Zinovy Malkin\footnote{Current affiliation: Pulkovo Observatory, St.~Petersburg, Russia} \\ Institute of Applied Astronomy, St.~Petersburg, Russia}
\date{\vspace{-10mm}}
\maketitle

\begin{abstract}
To estimate real accuracy of EOP prediction real-time predictions made by
the IERS Subbureau for Rapid Service and Prediction (USNO) and at the Institute of Applied Astronomy (IAA)
EOP Service are analyzed.
Methods of a priory estimate of accuracy of prediction are discussed.
\end{abstract}

\section{Introduction}

Forecast of various time series has been widely using
in many fields of science and practice and
many methods have been advanced for prediction of time series.
But common problem of each method is a priori estimate of its accuracy.

Common practice is to take a truncated series (reference series)
ended in the past, investigate its statistical parameters,
build prediction and
compare it with existing continuation of series under investigation.
Using moving shift of reference series one can collect needed statistics
and obtain estimate of accuracy of used method depending on
length of prediction.
After that obtained accuracy is assigned to real predictions.

In most of predicted time series last observed point (epoch)
preceding the first predicted one has its final value and
is not subject of refinement in future (e.g., number
of sunspots for some epoch).
It is not the case for EOP.  All real EOP predictions are
based on operational solution that may differ substantially
from final EOP values that comes usually in one-two months.

\begin{figure}[ht]
\epsfxsize=100mm
\centerline{\epsfbox{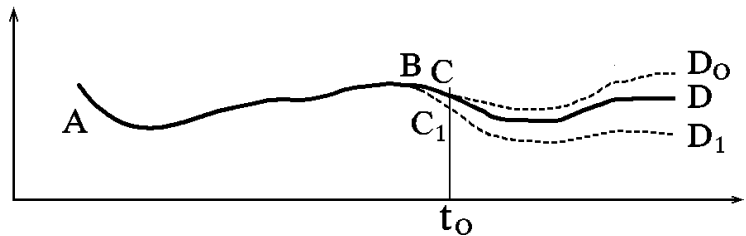}}
\caption{Influence of errors in operational EOP on prediction.}
\label{fig:pred1}
\end{figure}

As an illustration of the foregoing, let's consider Fig.~1.
In this figure
$t_0$ -- the epoch of beginning of prediction,
$A-B-C_1$ \ -- operational EOP series used for computation of prediction,
$C_1-D_1$ \ -- prediction computed in real time (at date $t_0$),
$A-B-C-D$ \ -- final EOP series,
$C-D_0$ \ -- prediction that would be computed from the final series
if it were available at date $t_0$.
Upon this circumstance, accuracy and other statistical parameters
of operational EOP series may differ substantially from
ones for final series that are commonly used for a priori
estimate of prediction accuracy.
It means that estimates of prediction accuracy obtained by ``standard''
method may be far from reality and some its modification (or at least
investigation) is desirable.

\section{A priori estimate of real accuracy of EOP prediction}

Evidently, the most simple way to make a priory estimate of
accuracy of prediction is a fictive disturbances of one or more
last points of reference interval to investigate reaction of
given method of prediction on errors at the last observed epoch(s).
This test was realized in (Malkin \& Skurikhina 1996).
Two kind of fictive errors was applied to real observed points:

{\em Test 1}: \ The value of 1 mas was added to (or subtract from)
   the C04 value corresponding to the last observed epoch.

{\em Test 2}: \ The values of 0.5, 1.0, 1.5 mas were added to (or subtract
   from) the C04 value corresponding to the three last observed epoch.

This test was used only for ARIMA
method because its influence on the extrapolation of trend-harmonics
model (e.g., McCarthy \& Luzum 1991)
can be easily foreseen without special calculations.
Typical differences between predictions of real and distorted
C04 series are presented in Table~\ref{tab:pred_xy_dist}.
One can see that serious degradation of accuracy may occur when
ARIMA method is used for erroneous observed EOP values.
It should be mentioned
that this effect practically linearly depends on the value of error.
Analogous results were obtained for prediction of UT
(Malkin \& Skurikhina 1996).

\begin{table}[b]
\caption{Influence of errors in the last values on prediction results.}
\begin{center}
\begin{tabular}{crrrrrrrr}
Test&\multicolumn{8}{c}{Length of prediction, days} \\
& \multicolumn{1}{c}{1} & \multicolumn{1}{c}{3} &
\multicolumn{1}{c}{5} & \multicolumn{1}{c}{10} &
\multicolumn{1}{c}{20} & \multicolumn{1}{c}{30} &
\multicolumn{1}{c}{60} & \multicolumn{1}{c}{90} \\
\hline
$1$ & 2.6 & 5.9 & 7.1 & 6.8 & 6.2 & 4.2 & 2.3 & 1.9 \\
$2$ & 1.2 & 2.2 & 2.2 & 2.1 & 1.9 & 1.7 & 1.4 & 1.2 \\
\end{tabular}
\end{center}
\label{tab:pred_xy_dist}
\end{table}

It is clear that proposed test can be only useful for investigation of
sensitivity of a given method of prediction to errors of operational EOP.
To estimate a priori errors in real predictions by this method
we need (at least) to know real errors in operational EOP series
and their statistical parameters.

\section{Assessment of errors in real predictions}

For this investigation we use two series: real predictions produced at IAA and USNO.
These two centers use different methods for
short-time prediction (ARIMA at IAA and trend-harmonics model
at USNO), whereas methods of long-time prediction are
similar (McCarthy \& Luzum 1991, Malkin \& Skurikhina 1996).
The data for the period from February 1998 through August 1999.
To reduce amount of data only predictions computed on Thursdays were used.

Results are presented in Table~\ref{tab:pred1},
where column ``IAA-R'' contains rms errors in real predictions
computed at the IAA,
column ``NEOS'' contains rms errors in predictions computed at the USNO
and published in the IERS Bulletin A.
Column ``IAA-A'' contains predictions re-computed a posteriori, in at least three months after the first day of prediction.
Column ``N'' contains number of used predictions.

\begin{table}
\caption{RMS errors in predictions (C04 for IAA).}
\begin{center}
\begin{tabular}{cccccccc}
Length of & \multicolumn{2}{c}{IAA-R} & \multicolumn{2}{c}{NEOS} &
\multicolumn{2}{c}{IAA-A} & N \\
prediction  & Pole & UT & Pole & UT & Pole & UT & \\
            & mas  & ms & mas  & ms & mas  & ms & \\
\hline
   0 &   0.29 &   0.14 &   0.19 &   0.06 &   0.03 &   0.00 &  76  \\
   1 &   0.59 &   0.18 &   0.48 &   0.10 &   0.23 &   0.04 &  76  \\
   2 &   0.98 &   0.27 &   0.85 &   0.17 &   0.63 &   0.13 &  76  \\
   3 &   1.42 &   0.35 &   1.25 &   0.26 &   1.06 &   0.23 &  76  \\
   5 &   2.21 &   0.59 &   1.98 &   0.50 &   1.88 &   0.44 &  76  \\
   7 &   2.85 &   0.87 &   2.56 &   0.83 &   2.60 &   0.70 &  75  \\
  10 &   3.82 &   1.44 &   3.50 &   1.38 &   3.63 &   1.20 &  75  \\
  30 &   9.53 &   5.23 &   9.59 &   5.05 &   9.56 &   5.04 &  73  \\
  60 &  15.1 &  10.9 &  15.8 &  11.1 &  15.3 &  10.3 &  69  \\
 120 &  21.4 &  25.7 &  25.4 &  28.3 &  21.6 &  25.2 &  60  \\
\end{tabular}
\end{center}
\label{tab:pred1}
%
\caption{RMS errors in predictions (NEOS for IAA).}
\begin{center}
\begin{tabular}{cccccccc}
Length of & \multicolumn{2}{c}{IAA-R} & \multicolumn{2}{c}{NEOS} &
\multicolumn{2}{c}{IAA-A} & N \\
prediction  & Pole & UT & Pole & UT & Pole & UT & \\
            & mas  & ms & mas  & ms & mas  & ms & \\
\hline
   0  &    0.18  &  0.05 &   0.18 &   0.05  &   0.03 &   0.01 & 48  \\
   1  &    0.38  &  0.07 &   0.46 &   0.10  &   0.25 &   0.03 & 48  \\
   2  &    0.73  &  0.13 &   0.82 &   0.17  &   0.62 &   0.09 & 48  \\
   3  &    1.16  &  0.21 &   1.21 &   0.27  &   1.04 &   0.15 & 48  \\
   5  &    1.89  &  0.42 &   1.86 &   0.53  &   1.89 &   0.31 & 48  \\
   7  &    2.54  &  0.68 &   2.40 &   0.87  &   2.54 &   0.58 & 47  \\
  10  &    3.38  &  1.17 &   3.19 &   1.39  &   3.42 &   1.11 & 47  \\
  30  &    8.15  &  5.01 &   7.54 &   4.83  &   8.18 &   5.07 & 45  \\
  60  &   11.8  & 11.2 &  11.1 &  11.1  &  11.9 &  11.2 & 41  \\
 120  &   16.2  & 24.4 &  19.3 &  24.8  &  16.3 &  24.5 & 32  \\
\end{tabular}
\end{center}
\label{tab:pred2}
\end{table}

Table~\ref{tab:pred1} shows that accuracy of predictions computed
at the IAA and the USNO is approximately the same except
the beginning of interval. Since short-time prediction is
the most interesting to users, that is worthwhile to investigate
it more carefully.
Evidently accuracy of prediction depends on predicted series.
IAA predictions are being made using EOP(IERS)C04 series
as reference, whereas USNO uses NEOS series.  If there is
substantial difference in accuracy of last epochs of these
two operational series it may cause difference in accuracy
of predictions.

To estimate accuracy of operational solutions
C04 and NEOS we have included in Table~\ref{tab:pred1}
the first line with length of prediction equal to zero.
This line contains, in fact, rms error in last reported epochs
of operational C04 series in columns related
to IAA predictions and error in the last reported epoch
of the NEOS series in column related to USNO predictions.
Comparison of these values shows that NEOS operational solution is more accurate than C04 one.
Comparison of predictions IAA-R and IAA-A shows that
accuracy of real predictions differs substantially from a priori
estimates for short-time prediction, whereas a priory
estimate for long-time prediction is adequate to reality.

Taking into account the difference in accuracy of operational
series used for prediction at the IAA and the USNO we tried
to perform another, more rigorous test to compare methods used
at these institutions.
For this purpose we compute prediction
of NEOS series using IAA method.
In parallel, we computed a posteriori predictions
in the same way as above but using NEOS series as reference, too.
Since we began to collect these predictions only
in October 1998, statistics for this test is more poor than
for previous one.  Table~\ref{tab:pred2} contains results
of this test. Notations are the same as in Table~\ref{tab:pred1}.

Using results of the last comparison we can conclude that accuracy
of predictions computed in both centers are approximately the same.
More detailed investigation requires more predictions involved in
statistics.
Again, one can see that a posteriori predictions
do not provide adequate estimate of accuracy of short-time prediction.

Another important index of quality of prediction is maximal error
in predicted EOP values that provide ``guaranteed'' error needed
in some practical applications.  Tables \ref{tab:pred1m} and
\ref{tab:pred2m} contains maximal errors in predictions.
They are analogous to Tables \ref{tab:pred1} and \ref{tab:pred2}.
Again, Table \ref{tab:pred1m} contents results for the period
from February 1998 through August 1999
and Table \ref{tab:pred2m} -- from October 1998 through August 1999.

\begin{table}
\caption{Maximal errors in predictions (C04 for IAA).}
\begin{center}
\begin{tabular}{cccccccc}
Length of & \multicolumn{2}{c}{IAA-R} & \multicolumn{2}{c}{NEOS} &
\multicolumn{2}{c}{IAA-A} & N \\
prediction  & Pole & UT & Pole & UT & Pole & UT & \\
            & mas  & ms & mas  & ms & mas  & ms & \\
\hline
   0  &   1.09 &   0.40 &    0.63 &   0.18  &    0.21 &   0.01 &  76  \\
   1  &   2.82 &   0.51 &    1.34 &   0.29  &    0.77 &   0.11 &  76  \\
   2  &   3.96 &   1.02 &    3.00 &   0.52  &    1.65 &   0.34 &  76  \\
   3  &   5.39 &   1.09 &    4.08 &   0.87  &    3.42 &   0.50 &  76  \\
   5  &   8.31 &   1.83 &    6.34 &   1.78  &    6.54 &   1.07 &  76  \\
   7  &  10.1 &   2.70 &    7.65 &   2.72  &    8.37 &   2.09 &  75  \\
  10  &  11.7 &   4.33 &    9.36 &   4.40  &   10.6 &   3.65 &  75  \\
  30  &  23.2 &  14.4 &   23.0 &  12.1  &   23.6 &  14.1 &  73  \\
  60  &  35.1 &  20.2 &   38.4 &  20.4  &   34.2 &  20.3 &  69  \\
 120  &  47.5 &  47.0 &   60.1 &  48.8  &   47.4 &  46.5 &  60  \\
\end{tabular}
\end{center}
\label{tab:pred1m}
%
\caption{Maximal errors in predictions (NEOS for IAA).}
\begin{center}
\begin{tabular}{cccccccc}
Length of & \multicolumn{2}{c}{IAA-R} & \multicolumn{2}{c}{NEOS} &
\multicolumn{2}{c}{IAA-A} & N \\
prediction  & Pole & UT & Pole & UT & Pole & UT & \\
            & mas  & ms & mas  & ms & mas  & ms & \\
\hline
   0  &   0.53 &   0.13 &   0.53 &   0.13   &   0.10 &   0.03 &   48  \\
   1  &   1.35 &   0.28 &   1.34 &   0.29   &   0.78 &   0.07 &   48  \\
   2  &   3.06 &   0.46 &   3.00 &   0.52   &   1.61 &   0.44 &   48  \\
   3  &   4.32 &   0.73 &   4.08 &   0.87   &   3.10 &   0.35 &   48  \\
   5  &   5.65 &   1.44 &   5.19 &   1.78   &   6.44 &   0.86 &   48  \\
   7  &   8.31 &   2.22 &   6.56 &   2.72   &   8.33 &   1.86 &   47  \\
  10  &  10.8 &   3.78 &   9.36 &   4.40   &  10.9 &   3.42 &   47  \\
  30  &  23.3 &  13.4 &  21.2 &  10.1   &  23.5 &  13.9 &   45  \\
  60  &  30.9 &  18.8 &  25.6 &  20.2   &  30.9 &  19.3 &   41  \\
 120  &  41.2 &  35.7 &  35.8 &  39.1   &  41.2 &  35.3 &   32  \\
\end{tabular}
\end{center}
\label{tab:pred2m}
\end{table}

\section{Conclusion}

In this paper we have attempted to estimate real accuracy of
predictions of EOP using real-time predictions made at the IAA and
the USNO has been also made.
Although collected statistics is too poor to make more or less final
conclusions, we can state that:
\begin{itemize}
\item Estimate of accuracy of prediction based on use of old data
    is not adequate to accuracy of real-time prediction, especially
    for short-time prediction.  A modification of commonly used method
    of a priori estimate of accuracy, e.g., proposed in
    (Malkin \& Skurikhina 1996) can give more realistic estimates.
\item Accuracy of methods of prediction of EOP used at the IAA and the
    USNO is approximately the same.  More detailed conclusion can be
    made only after collecting supplement statistics.
\item Estimate of both RMS and maximal errors in prediction
    is very useful for potential users.  It seems reasonable to
    provide such estimates for IERS and other prediction series.
\end{itemize}

\bigskip
\noindent{\Large\bf References}
\bigskip
\leftskip=\parindent
\parindent=-\leftskip

Malkin, Z., \& Skurikhina, E. 1996,  On Prediction of EOP, Comm. IAA, No 93. (arXiv:0910.3336)

McCarthy, D.D., \& Luzum, B.J. 1991, Prediction of Earth Orientation, Bull. Geod., v.~65, 18--21.

\end{document}